\documentclass[twocolumn,twoside,slac_two]{revtex4}
\usepackage{graphicx}
\usepackage{fancyhdr}
\def\babar{\mbox{\slshape B\kern-0.1em{A}\kern-0.1em
    B\kern-0.1em{A\kern-0.2em R}}}
\newcommand{\bzb}{\bar{B}^0}
\newcommand{\Dt}{\Delta t}
\newcommand{\Dm}{\Delta m}
\newcommand{\aeff}{\alpha_{\rm eff}}
\newcommand{\bhh}{B \to h h}
\newcommand{\bpp}{B \to \pi \pi}
\newcommand{\bpppm}{B^0 \to \pi^+ \pi^-}
\newcommand{\bpzpz}{B^0 \to \pi^0 \pi^0}
\newcommand{\bpppz}{B^+ \to \pi^+ \pi^0}
\newcommand{\brp}{B \to \rho \pi}
\newcommand{\brppm}{B^0 \to \rho^{\pm} \pi^{\mp}}
\newcommand{\bppp}{B^0 \to \pi^+ \pi^- \pi^0}
\newcommand{\brr}{B \to \rho \rho}
\newcommand{\brprm}{B^0 \to \rho^+ \rho^-}
\newcommand{\brprz}{B^+ \to \rho^+ \rho^0}
\newcommand{\brzrz}{B^0 \to \rho^0 \rho^0}
\newcommand{\pppm}{\pi^+ \pi^-}
\newcommand{\pppz}{\pi^+ \pi^0}
\newcommand{\pzpz}{\pi^0 \pi^0}
\newcommand{\rprm}{\rho^+ \rho^-}
\newcommand{\rprz}{\rho^+ \rho^0}
\newcommand{\rzrz}{\rho^0 \rho^0}
\newcommand{\bkr}{B^+ \to K^{*0} \rho^+}
\newcommand{\ksr}{K^{*0} \rho^+}
\newcommand{\bap}{B \to a_1 \pi}
\newcommand{\bappm}{B^0 \to a_1^{\pm} \pi^{\mp}}
\newcommand{\ap}{a_1 \pi}

\begin{document}

\title{Measurements of the angle $\alpha$ ($\phi_2$) at B factories}

%

\author{G. Vasseur}
\affiliation{CEA, Irfu, SPP, Centre de Saclay, F-91191 Gif sur Yvette, France}

\begin{abstract}
The measurements of the angle $\alpha$ ($\phi_2$) of the unitarity triangle 
at the B factories are reviewed.
The value of $\alpha$ determined by combining the results obtained 
in the $B \to \pi \pi$, $B \to \rho \pi$, and $B \to \rho \rho$ modes 
by both the \babar\ and Belle experiments is 
$(87.5^{+6.2}_{-5.3}){\ensuremath{^\circ}}$.
\end{abstract}

\maketitle

\thispagestyle{fancy}


\section{Introduction}

\subsection{The angle $\alpha$}
\begin{figure}[htb]
\centering
\includegraphics[width=80mm]{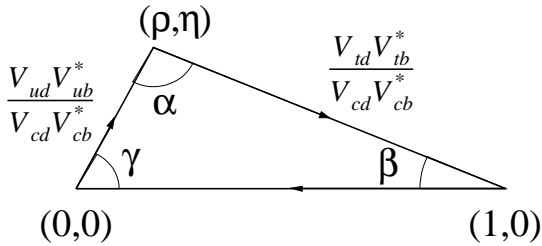}
\caption{The unitariry triangle.} \label{ut}
\end{figure}

In the Standard Model (SM), $CP$ violation in the quark sector is explained 
through the complex Cabibbo-Kobayashi-Maskawa (CKM) quark-mixing 
matrix~\cite{ckm}.
One relation due the unitarity of the CKM matrix,
$V_{ud}V^*_{ub}+V_{cd}V^*_{cb}+V_{td}V^*_{tb}=0$,
is represented graphically in the complex plane 
as the unitarity triangle, shown in Figure~\ref{ut}.

The goal of the \babar\ and Belle experiments is to test the SM explanation 
of $CP$ violation in overconstraining the unitarity triangle,
by measuring its sides and its three angles $\alpha$, $\beta$, and $\gamma$
(also called $\phi_2$, $\phi_1$, and $\phi_3$ respectively).
The angle $\alpha = {\rm arg}(-\frac{V_{td}V^*_{tb}}{V_{ud}V^*_{ub}})$ brings
into play the two elements of the CKM matrix 
which are non real at lowest order:
$V_{ub}$, involved in decays proceeding via a $b$ to $u$ transtion, 
whose phase is $-\gamma$,
and $V_{td}$, involved in $B^0 \bzb$ mixing, whose phase is $-\beta$.
So $\alpha = \pi - \beta - \gamma$ can be measured 
from $CP$ violating asymmetries in the interference between 
mixing and decay in charmless decays of the neutral $B$ mesons,
such as $\bpppm$, $\brprm$, $\bppp$, and $\bappm$.

\subsection{Time-dependent $CP$ asymmetry}
A $B^0$ meson can decay into a $CP$ eigenstate $f$ either directly 
or after having oscillated to a $\bzb$ meson which then decays to $f$.
The amplitudes associated to the two processes are respectively $A_f$ and
$\frac{q}{p} \bar{A}_f$ where $\frac{q}{p} \sim e^{-2i\beta}$
accounts for $B^0 \bzb$ mixing.
The resulting $CP$ violation asymmetry as a function of the proper time
difference $\Dt$ can be expressed as:

\begin{eqnarray}
a_f (\Dt) & = & \frac {\Gamma_{\bzb \to f}(\Dt) - \Gamma_{B^0 \to f}(\Dt)} {\Gamma_{\bzb \to f}(\Dt) + \Gamma_{B^0 \to f}(\Dt)}  \nonumber \\
& = & S \sin (\Dm \Dt) -  C \cos (\Dm \Dt)
\label{eq-asym}
\end{eqnarray}

\noindent where $\Dm$ is the mass difference 
between the two neutral $B$ mass eigenstates.
The coefficients $C = \frac{1-|\lambda_f|^2}{1+|\lambda_f|^2}$ and 
$S = \frac{2\Im(\lambda_f)}{1+|\lambda_f|^2}$ 
are functions of the ratio of the amplitudes 
with and without mixing $\lambda_f = \frac{q}{p} \frac{\bar{A}_f}{A_f}$.
$C$ (or ${\cal A} = -C$) measures direct $CP$ violation 
while $S$ measures $CP$ violation in the interference between decay and mixing.

In the simple case of a charmless $B^0$ decay 
involving only the $b$ to $u$ tree diagram amplitude, with weak phase
$\gamma$, we have $\lambda_f = e^{-2i\beta}e^{-2i\gamma} = e^{2i\alpha}$,
resulting in $C = 0$ and $S = \sin 2 \alpha$.
A time-dependent analysis of the $CP$ asymmetry in this mode would give 
a direct measurement of $\alpha$.
However other diagrams are involved in charmless decays and 
in particular the one loop gluonic penguin diagrams.
As the dominant gluonic penguin diagram does not carry the same weak phase
as the tree diagram, the extraction of $\alpha$ becomes more complex.
$C$ is no longer equal to 0 and $S = \sqrt{1-C^2} \sin 2 \aeff$ does not
any more measure $\alpha$ but an effective value $\aeff$.

\subsection{The isospin analysis}
The isospin analysis~\cite{gronau_london}, based on $SU(2)$ symmetry,
allows the extraction of the true value of $\alpha$.
It can be applied both to the $\bpp$ and $\brr$ modes and
uses all the $B$ decay modes to $hh$ ($h=\pi$ or $\rho$) to determine
the difference $\alpha - \aeff$.
Tree and penguin contributions to $\bhh$ decays are summarized in
Table~\ref{treepeng}. 
Since the tree amplitude in the $B^0 \to h^0 h^0$ decays is color suppressed,
the branching ratio of these modes is expected to be small.
Also the isospin conservation rules exclude gluonic penguin transitions
for $h^+h^0$ final states.

\begin{table}[htb]
\begin{center}
\caption{Diagrams contributing to $\bhh$ decays.}
\begin{tabular}{|c|c|c|}
\hline
Mode & Tree & Gluonic penguin \\
\hline
$ h^+h^-$ & Color-allowed     & Present    \\
$ h^0h^0$ & Color-suppressed  & Present    \\ 
$ h^+h^0$ & Color-allowed     & Forbidden  \\
\hline
\end{tabular}
\label{treepeng}
\end{center}
\end{table}

The $SU(2)$ isospin symmetry relates the amplitudes of all the $\bhh$ modes:  
$A^{+-}=A(B^0 \to h^+ h^-)$, $A^{+0}=A(B^+ \to h^+ h^0)$,
$A^{00}=A(B^0 \to h^0 h^0)$, $\widetilde{A}^{+-}=A(\bzb \to h^+ h^-)$, 
$\widetilde{A}^{-0}=A(B^- \to h^- h^0)$, and
$\widetilde{A}^{00}=A(\bzb \to h^0 h^0)$. 
Neglecting electroweak penguins and other $SU(2)$-breaking effects, 
we obtain the isospin relations:

\begin{equation}
\frac{A^{+-}}{\sqrt{2}} + A^{00} = A^{+0} = \widetilde{A}^{-0} = 
\frac{\widetilde{A}^{+-}}{\sqrt{2}} + \widetilde{A}^{00}
\end{equation}

\noindent
where we used the fact that the amplitude of the pure tree 
$B^+ \to h^+ h^0$ mode is equal to the one of its charge conjugate process.

\begin{figure}[htb]
\centering
\includegraphics[width=80mm]{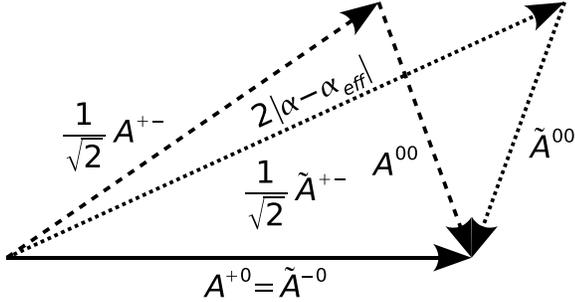}
\caption{Graphical representation of the isospin relations.} 
\label{fig:isospin}
\end{figure}

These relations are represented graphically as two triangles 
with a common base, shown in Figure \ref{fig:isospin}. 
Measuring the lengths of the sides of the two triangles, 
related to the branching ratios of the various modes, 
constrains $\alpha -\aeff$.  
As both triangles have two possible orientations,
up and down, the isospin method has a four-fold ambiguity,
which comes in addition to the two-fold ambiguity related to the fact
that only the sine of $2 \aeff$ is measured 
in the time-dependent analysis of $B^0 \to h^+ h^-$.

\subsection{Analyses overview}
The \babar\ and Belle detectors at the PEP-II and KEK-B colliders respectively 
are described in details elsewhere~\cite{detector}.

At the $\Upsilon(4S)$ resonance,
$B \bar{B}$ pairs are produced in a coherent state.  
The $CP$ asymmetry is measured as a function of the difference 
between the decay times of the two $B$ mesons $\Dt$,
which in an energy asymmetric machine is obtained from the measured
distance along the beam axis between the two vertices $\Delta z$
according to $\Delta t = \Delta z / \beta \gamma c$.
Here $c$ is the speed of light and 
$\beta \gamma$, equal to 0.56 in \babar\ and 0.425 in Belle, 
is the Lorentz boost of the $\Upsilon(4S)$.

One of the $B$ mesons is fully reconstructed into the $B$ decay of interest 
($\pi \pi$, $\rho \pi$, $\rho \rho$, or $a_1 \pi$), 
while the other $B$ meson is used to tag its flavor at production time. 
Tagging combines different techniques including the use
of semileptonic decays and secondary kaons. 

To select the signal, hadron identification is used to separate pions
from kaons.
The beam energy constrained $B$-meson mass
$m_{B}=\sqrt{E^{2}_{\rm beam}-p_B^{2}}$ 
and the energy difference $\Delta E=E_B-E_{\rm beam}$ 
are powerful kinematic discriminating variables, 
peaking respectively for the signal at the $B$-meson mass and zero.  
$E_{\rm beam}$ is the beam energy.
$E_B$ and $p_B$ are the energy and momentum
of the reconstructed candidate, all evaluated in the $\Upsilon(4S)$ rest frame.
 
The largest background consists of $q\bar q$ ($q=u,d,s,c$) continuum events.
The $B\bar B$ events look spherical while the continuum events
are jet-like, so event shape variables are also used for separation. 

The following results are based on multi-variable maximum likelihood analyses.

\section{Results from $B\to \pi \pi$ decays}

The time-dependent $CP$ asymmetry of the $B^0 \to \pi^+ \pi^-$ decays has been 
studied by Belle using $535 \times 10^6~B\bar B$ pairs \cite{pipi_belle} and
by \babar\ using $383 \times 10^6~B\bar B$ pairs \cite{pipi_babar}. 
Fits that include respectively $1464 \pm 65$ signal events
for Belle and $1139 \pm 49$ for \babar\ give measurements of $S_{\pppm}$
and $C_{\pppm}$, summarized in Table~\ref{SCpp}.
Both experiments have observed $CP$ violation 
in the interference between decay and mixing in the $\bpppm$ decays
as the $S_{\pppm}$ parameter is found different from 0 
with a significance of 5.3~$\sigma$ for Belle and 5.1~$\sigma$ for \babar\
with a perfect agreement between the two experiments.
In addition Belle sees a 5.5~$\sigma$ effect also for the direct $CP$
violation parameter $C_{\pppm}$, which is different from the \babar\ result,
at the level of 2.1~$\sigma$, but not inconsistent with it.

\begin{table*}[tb]
\begin{center}
\caption{Measurements of $CP$ parameters and branching fractions 
in the $\bpp$ modes.}
\begin{tabular}{|l|c|c|c|}
\hline
 & \babar\ & Belle & World average \\
\hline
$S_{\pppm}$   & $-0.60 \pm 0.11 \pm 0.03$ & $-0.61 \pm 0.10 \pm 0.04$ &
 $-0.61 \pm 0.08$ \\
$C_{\pppm}$   & $-0.21 \pm 0.09 \pm 0.02$ & $-0.55 \pm 0.08 \pm 0.05$ &
 $-0.38 \pm 0.07$ \\
\hline
${\cal A}_{\pppz}$ & $0.03 \pm 0.08 \pm 0.01$ & $0.07 \pm 0.06 \pm 0.01$ &
 $0.06 \pm 0.05$ \\
${\cal A}_{\pzpz}$ & $0.5 \pm 0.4 \pm 0.1$ & $0.4 \pm 0.7 \pm 0.1$ &
 $0.5 \pm 0.3$ \\
\hline
${\cal B}(\bpppm)$ & $(5.5 \pm 0.4 \pm 0.3) 10^{-6}$ & 
 $(5.1 \pm 0.2 \pm 0.2) 10^{-6}$ & $(5.2 \pm 0.2) 10^{-6}$ \\
${\cal B}(\bpppz)$ & $(5.0 \pm 0.5 \pm 0.3) 10^{-6}$ &
 $(6.5 \pm 0.4 \pm 0.5) 10^{-6}$ & $(5.6 \pm 0.4) 10^{-6}$ \\
${\cal B}(\bpzpz)$ & $(1.5 \pm 0.3 \pm 0.1) 10^{-6}$ &
 $(1.1 \pm 0.3 \pm 0.1) 10^{-6}$ & $(1.3 \pm 0.2) 10^{-6}$ \\
\hline
\end{tabular}
\label{SCpp}
\end{center}
\end{table*}

All the $\bpp$ modes, which are needed to perform the isospin analysis,
have been measured by \babar\ and Belle.
The measurements from \babar\ are based on $227 \times 10^6~B\bar B$ pairs 
for $\bpppm$~\cite{pippim_babar} and on $383 \times 10^6~B\bar B$ pairs 
for $\bpppz$ and $\bpzpz$~\cite{pipiz_babar}, while those from Belle use
$449 \times 10^6~B\bar B$ pairs for $\bpppm$ and $\bpppz$~\cite{pippi_belle}
and $535 \times 10^6~B\bar B$ pairs for $\bpzpz$~\cite{pizpiz_belle} and
for the $CP$ asymmetry in $\bpppz$~\cite{cpasym_belle}.
The results and the world averages~\cite{hfag} 
for the branching ratios over the charge conjugate processes,
as well as the time-integrated charge asymmetries ${\cal A}$ are given 
in Table~\ref{SCpp}.
The branching ratio of $B^0 \to \pi^0 \pi^0$,
whose tree diagram is color supressed, is relatively large,
about one fourth of that of $\bpppm$,
thus indicating a large penguin contamination in $\bpp$.
The charge asymmetry in $\bpppz$ is compatible with 0 as expected.

\begin{figure}[htb]
\centering
\includegraphics[width=80mm]{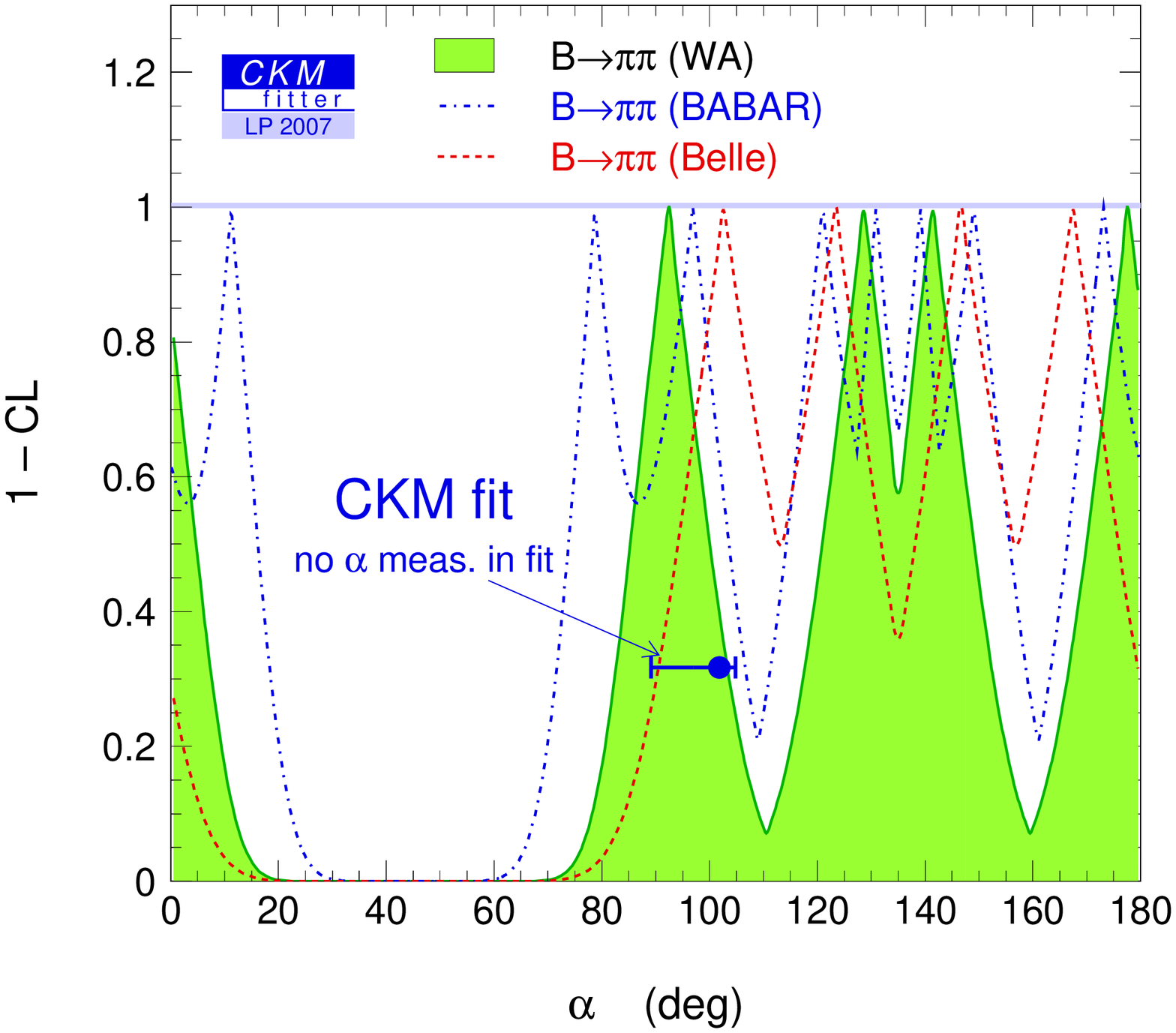}
\caption{Constraint on $\alpha$ from the isospin analysis in the $\bpp$ modes.
The dot-dashed (dashed) curve uses \babar\ (Belle) measurements only, 
while the hatched region gives the combined constraint.}
\label{alpha_pipi}
\end{figure}

Using the six observables (the branching fraction of the three $\bpp$ modes,
$S_{\pppm}$, $C_{\pppm}$, and ${\cal A}_{\pzpz}$) 
an isospin analysis is performed to determine the six unknown parameters
and in particular $\alpha$.
The constraint on $\alpha$ is illustrated by the confidence level plot 
from the CKMfitter group~\cite{ckm_fitter} in Figure~\ref{alpha_pipi}.
The already mentionned 8-fold ambiguity in the determination of $\alpha$
corresponds to the eight peaks which can be seen on this plot.
Choosing the peak consistent with the other CKM measurements,
the combined determination of $\alpha$ from the $\bpp$ modes is 
$(92^{+11}_{-10}){\ensuremath{^\circ}}$.
The range of values between $14{\ensuremath{^\circ}}$ and 
$76{\ensuremath{^\circ}}$ is excluded at 95\% C.L.

\section{Results from $B\to \rho \rho$ decays}

$\brr$ analyses are experimentally more challenging than the $\bpp$ analyses
as the final states consist of four pions, including two $\pi^0$ 
for the $\rprm$ mode. 
The wide $\rho$ resonances also result in more background. 
Finally these vector-vector modes are not $CP$ eigenstates. 

But as the $\brprm$ mode is almost $100$\% longitudinally polarized,
an analysis of the sole longitudinal $CP$-even component is adequate.
Besides the branching ratio for $\brprm$ is about five times larger than 
for $\bpppm$, and the ratio of penguin over tree amplitude is smaller.
Thus this mode turns out to be better  for constraining $\alpha$.

A similar analysis to that for $\bpppm$  is performed for the $\brprm$ mode,
with the reconstructed masses of the two $\rho$ mesons
as well as their helicity angles as additional observables, and the
fraction $f_L$ of longitudinal polarization as an additional
parameter.  The decay rate as a function of $f_L$ and the
helicity angles $\theta_{1,2}$ is:
$\frac{d\Gamma}{d\theta _1 d\theta _2} \propto
\frac{1-f_L}{4}\sin^2\theta_1\sin^2\theta_2+f_L\cos^2\theta_1\cos^2\theta_2$.

Using respectively $576 \pm 53$ signal events 
from a sample of $535\times 10^6$ $B\bar B$ pairs~\cite{rhorho_belle}
and $729 \pm 60$ signal events 
from a sample of $383\times 10^6$ $B\bar B$ pairs~\cite{rhorho_babar},
the Belle and \babar\ experiments have measured the $CP$ parameters $S_{\rprm}$
and $C_{\rprm}$, given in Table~\ref{SCrr}.

\begin{table*}[tb]
\begin{center}
\caption{Measurements of $CP$ parameters, branching fractions, 
and fractions of longitudinal polarization in the $\brr$ modes.}
\begin{tabular}{|l|c|c|c|}
\hline
 & \babar\ & Belle & World average \\
\hline
$S_{\rprm}$   & $-0.17 \pm 0.20 \pm 0.06$ & $ 0.19 \pm 0.30 \pm 0.08$ &
 $-0.05 \pm 0.17$ \\
$C_{\rprm}$   & $ 0.01 \pm 0.15 \pm 0.06$ & $-0.16 \pm 0.21 \pm 0.08$ &
 $-0.06 \pm 0.13$ \\
\hline
${\cal A}_{\rprz}$ & $-0.12 \pm 0.13 \pm 0.10$ & $0.00 \pm 0.22 \pm 0.03$ &
 $-0.08 \pm 0.13$ \\
\hline
$C_{\rzrz}$ & $0.4 \pm 0.9 \pm 0.2$ & - & $0.4 \pm 0.9$ \\
$S_{\rzrz}$ & $0.5 \pm 0.9 \pm 0.2$ & - & $0.5 \pm 0.9$ \\
\hline
${\cal B}(\brprm)$ & $(25 \pm 2 \pm 4) 10^{-6}$ & 
 $(23 \pm 4 \pm 3) 10^{-6}$ & $(24 \pm 3) 10^{-6}$ \\
${\cal B}(\brprz)$ & $(17 \pm 2 \pm 2) 10^{-6}$ &
 $(32 \pm 7 ^{+4}_{-7}) 10^{-6}$ & $(18 \pm 3) 10^{-6}$ \\
${\cal B}(\brzrz)$ & $(0.8 \pm 0.3 \pm 0.2) 10^{-6}$ &
 $(0.4 \pm 0.4 \pm 0.2) 10^{-6}$ & $(0.7 \pm 0.3) 10^{-6}$ \\
\hline
$f_L^{\rprm}$   & $ 0.99 \pm 0.02 \pm 0.02$ & $0.94 \pm 0.04 \pm 0.03$ &
 $0.98 \pm 0.02$ \\
$f_L^{\rprz}$ & $0.90 \pm 0.04 \pm 0.03$ & $0.95 \pm 0.11 \pm 0.02$ &
 $0.91 \pm 0.04$ \\
$f_L^{\rzrz}$ & $0.70 \pm 0.14 \pm 0.05$ & - & $0.70 \pm 0.15$ \\
\hline
\end{tabular}
\label{SCrr}
\end{center}
\end{table*}

The other ingredients for the isospin analysis, the branching fractions,
fractions of longitudinal polarization and remaining $CP$ parameters, 
have been measured by Belle using $275 \times 10^6$ $B\bar B$ pairs 
for $\brprm$~\cite{rhoprhom_belle}, $85 \times 10^6$ $B\bar B$ pairs 
for $\brprz$~\cite{rhoprhoz_belle}, and $657 \times 10^6$ $B\bar B$ pairs 
for $\brzrz$~\cite{rhozrhoz_belle}, and by \babar\ 
using $383 \times 10^6$ $B\bar B$ pairs for $\brprm$~\cite{rhorho_babar},
$232 \times 10^6$ $B\bar B$ pairs for $\brprz$~\cite{rhoprhoz_babar},
and $427 \times 10^6$ $B\bar B$ pairs for $\brzrz$~\cite{rhozrhoz_babar}. 

The small value of the $\brzrz$ branching ratio compared to 
the one of the other channels shows that the penguin contributions are small 
in the $\brr$ modes.
The two experiments obtain somewhat different results for the $\brzrz$ mode,
though not inconsistent.
While Belle does not see any significant signal, \babar\ finds evidence
for this decay based on $85 \pm 28$ signal events with a significance 
of 3.6~$\sigma$ taking into account the systematics.
As the decay vertex of the $\rzrz$ final state can be reconstructed
in contrast to $\pzpz$, a time dependant analysis is possible,
leading to the measurement of $C_{\rzrz}$ and $S_{\rzrz}$.
\babar\ has demonstrated the feasibility of such an analysis and gives 
a first measurement of these two $CP$ parameters,
though with a large statistical error.

\begin{figure}[htb]
\centering
\includegraphics[width=80mm]{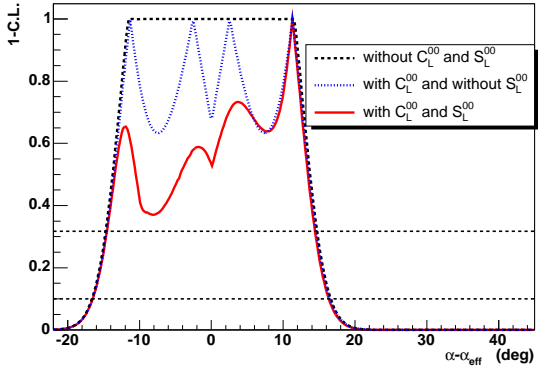}
\caption{Confidence level on $\alpha-\aeff$ obtained from 
the isospin analysis including the measurements of $C_{\rzrz}$ and $S_{\rzrz}$ 
(solid curve), the measurement of $C_{\rzrz}$ only (dotted curve) and 
none of them (dashed curve).
The horizontal dashed lines correspond to the $68\%$ (top) and $90\%$ 
(bottom) CL  intervals.}
\label{dalpha_rho0rho0}
\end{figure}

Figure~\ref{dalpha_rho0rho0} shows the impact of measuring these two parameters
for the isospin analysis in $\brr$ on the confidence level plot 
for $\alpha-\aeff$.
Without $C_{\rzrz}$ and $S_{\rzrz}$, we have only five observables
for six unknown parameters and the isospin analysis gives 
a plateau of degenerated values for $\alpha-\aeff$.
Adding $C_{\rzrz}$, we have now as many observables as unknown and find 
the four ambiguities of the isospin analysis.
If we also measure $S_{\rzrz}$, which gives seven observables for six unknown,
the isospin analysis is able to favor one single solution.
With the current statistics, the discrimination between the four solutions
is however limited.

\begin{figure}[htb]
\centering
\includegraphics[width=80mm]{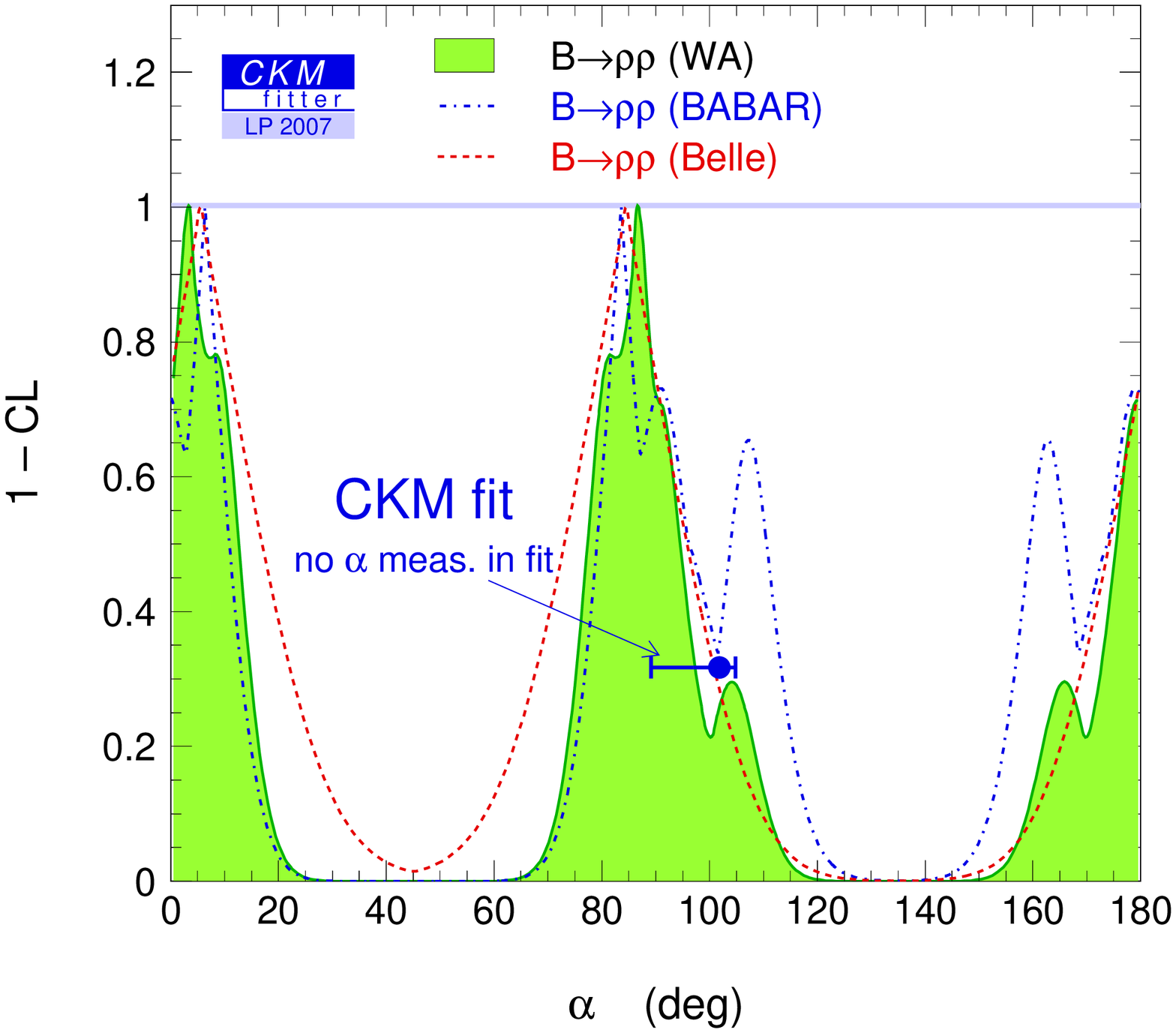}
\caption{Constraint on $\alpha$ from the isospin analysis in the $\brr$ modes.
The dot-dashed (dashed) curve uses \babar\ (Belle) measurements only, 
while the hatched region gives the combined constraint.}
\label{alpha_rhorho}
\end{figure}

Figure~\ref{alpha_rhorho} shows the outcome of the isospin analysis 
in the $\brr$ modes.
Choosing the solution consistent with the other CKM measurements,
the combined determination of $\alpha$ from the $\brr$ modes is 
$(87^{+10}_{-11}){\ensuremath{^\circ}}$.
The range of values between $20{\ensuremath{^\circ}}$ and 
$70{\ensuremath{^\circ}}$and the one between $113{\ensuremath{^\circ}}$ and 
$156{\ensuremath{^\circ}}$ are excluded at 95\% C.L.

\begin{table*}[tb]
\begin{center}
\caption{Measurements of the branching fraction, 
fraction of longitudinal polarization, and $CP$ parameter in the $\bkr$ modes.}
\begin{tabular}{|l|c|c|c|}
\hline
 & \babar\ & Belle & World average \\
\hline
${\cal B}(\bkr)$ & $(9.6 \pm 1.7 \pm 1.5) 10^{-6}$ & 
 $(8.9 \pm 1.7 \pm 1.2) 10^{-6}$ & $(9.2 \pm 1.5) 10^{-6}$ \\
$f_L^{\ksr}$   & $ 0.52 \pm 0.10 \pm 0.04$ 
& $0.43 \pm 0.11 ^{+0.05} _{-0.002}$ & $0.48 \pm 0.08$ \\
${\cal A}_{\ksr}$   & $-0.01 \pm 0.16 \pm 0.02$ & - & $-0.01 \pm 0.16$ \\
\hline
\end{tabular}
\label{kstrho}
\end{center}
\end{table*}

\begin{figure}[htb]
\centering
\includegraphics[width=80mm]{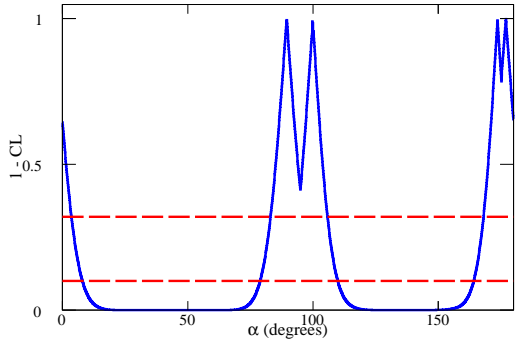}
\caption{Confidence level on $\alpha$ from the SU(3) analysis
of the $\brprm$ and $\bkr$ modes.
The horizontal dashed lines correspond to the $68\%$ (top) and $90\%$ 
(bottom) CL  intervals.}
\label{alpha_su3}
\end{figure}

An alternative approach to measure $\alpha$ is to use 
flavor SU(3) symmetry~\cite{beneke},
constraining the penguin contribution in $\brprm$ 
with the longitudinal part of the SU(3) partner mode: 
the pure penguin $\bkr$ channel.
The latter mode has been measured by Belle using 
$275 \times 10^6$ $B\bar B$ pairs ~\cite{kstrho_belle} and \babar\
using $232 \times 10^6$ $B\bar B$ pairs~\cite{kstrho_babar}.
This method has three unknowns: the ratio of penguin over tree amplitudes,
the relative phase between penguin and tree amplitudes and $\alpha$.
Taking SU(3) breaking effect into account, it gives a good constraint
on $\alpha$~\cite{rhorho_babar}: $83 < \alpha < 106{\ensuremath{^\circ}}$ 
at 68~\% C.L., illustrated in Figure~\ref{alpha_su3}.

\section {Results fom $B^0 \to \pi^+ \pi^- \pi^0$ Dalitz analysis.}

\begin{figure}[htb]
\centering
\includegraphics[width=60mm]{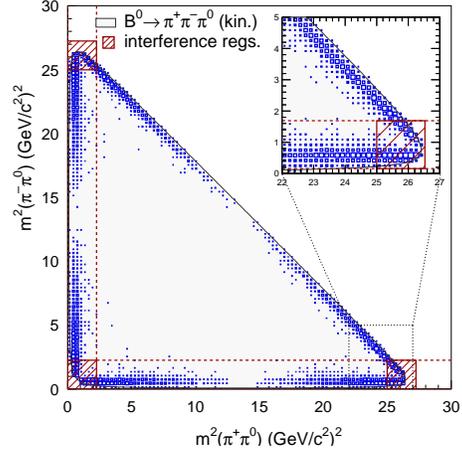}
\caption{Dalitz plot for Monte Carlo generated 
$B^0 \to \pi^+ \pi^- \pi^0$ events.
The main overlap regions between the $\rho$ bands are indicated 
by the hatched areas.}
\label{dalitz_rhopi}
\end{figure}

The $\brppm$ decay has no final $CP$ eigenstate like
$\pi^+\pi^-$ or $\rho^+\rho^-$. An isospin analysis would not
constrain sufficiently the many amplitudes of the $B$ decays to
$\rho^+\pi^-$, $\rho^-\pi^+$, $\rho^0\pi^0$, $\rho^+\pi^0$,
$\rho^0\pi^+$ and their charge conjugates.  A better approach
\cite{dalitz_3pi_theo} is based on the time-dependent analysis of the
$B^0 \to \pi^+ \pi^- \pi^0$ decay over the Dalitz plot, 
illustrated in Figure~\ref{dalitz_rhopi}, 
using the isospin symmetry as an additional constraint.  As this $B \to 3\pi$
decay is dominated by $\rho \pi$ resonances,
its amplitude is a function of well-known kinematic functions of the
Dalitz variables and of the $B^0 \to \rho \pi$ amplitudes,
themselves functions of $\alpha$ and tree and penguin contributions,
$A(B^0\to \rho^\kappa \pi^{-\kappa})=T^{\kappa}e^{-i\alpha}+P^{\kappa}$
with $\kappa$=$(0,+,-)$. 
Only the sign of the weak phase $\alpha$ is
changed when switching to the charge conjugate process.
The time-dependent $CP$ analysis of the $\bppp$ decay then provides 
enough constraints to extract $\alpha$ without discrete ambiguities
and the tree and penguin amplitudes. 

\begin{figure}[htb]
\centering
\includegraphics[width=80mm]{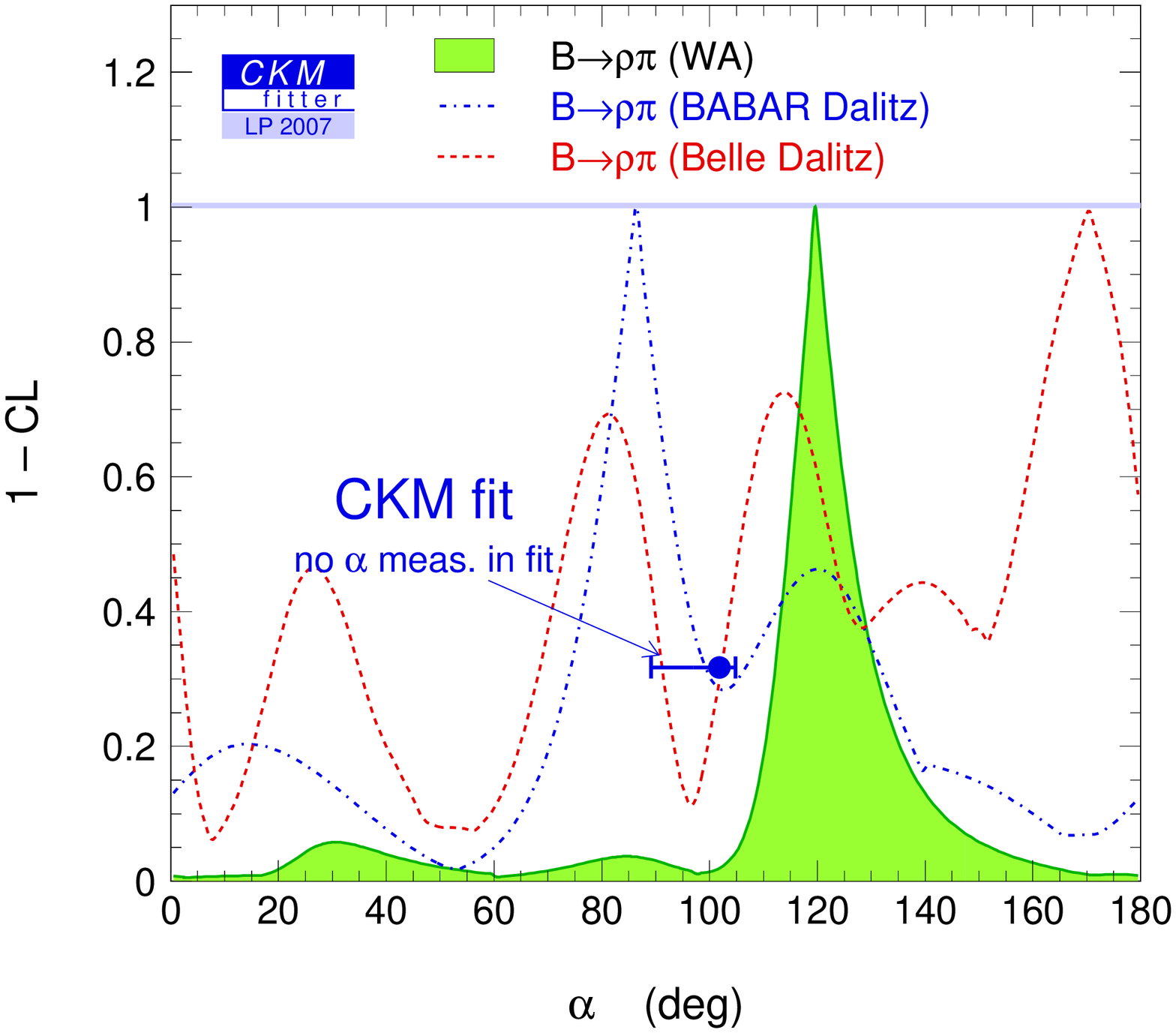}
\caption{Constraint on $\alpha$ from the Dalitz analysis in the $\brp$ modes.
The dot-dashed (dashed) curve uses \babar\ (Belle) measurements only, 
while the hatched region gives the combined constraint.}
\label{alpha_rhopi}
\end{figure}

Technically the analysis is based on a fit of 26 bilinear coefficients.
It has been performed by \babar\ 
using $375 \times 10^6~B\bar B$ pairs~\cite{rhopi_babar} and Belle
using $449 \times 10^6~B\bar B$ pairs~\cite{rhopi_belle}
corresponding to $2067 \pm 68$ and $971 \pm 42$ signal events respectively.
The resulting constraint on $\alpha$ is illustrated 
in Figure~\ref{alpha_rhopi}.
Combining the results of the two experiments is not just an average in $\alpha$
but a combination in the 26 experimentally measured coefficients,
which are correlated among each other.
The Dalitz analysis of the $\brp$ mode gives a combined value of $\alpha$
of $(120^{+11}_{-8}){\ensuremath{^\circ}}$.

\section{Study of $\bap$ decays}

The $\bap$ modes also allow the measurement of $\alpha$.
The $\bappm$ channel has a high branching fraction, 
$(33 \pm 4 \pm 3) 10^{-6}$ as measured by \babar\
using $218 \times 10^6~B\bar B$ pairs~\cite{a1piBF_babar} and 
$(30 \pm 3 \pm 5) 10^{-6}$ as measured by Belle
using $535 \times 10^6~B\bar B$ pairs~\cite{a1pi_belle}.
The two results are in good agreement and give a word average of
$(32 \pm 4) 10^{-6}$.

\begin{figure}[htb]
\centering
\includegraphics[width=80mm]{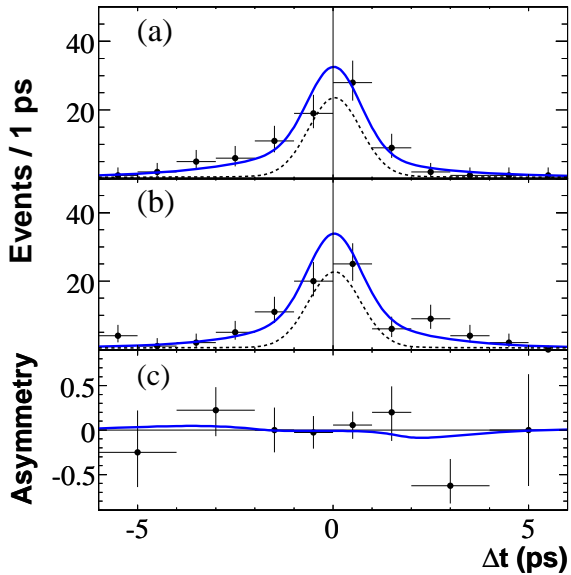}
\caption{Projections onto $\Dt$ of the data (points) for (a) $B^0$ and (b)
$\bzb$ tags, showing the fit function (solid line) and the backgrounf function
(dotted line), and (c) the asymmetry between $B^0$ and $\bzb$ tags.}
\label{time_a1pi}
\end{figure}

The $\bappm$ mode, like $\brppm$, is not a $CP$ eigenstate.
Using a quasi-two body approach, \babar\ has performed 
a time-dependent analysis of this mode 
based on $384 \times 10^6~B\bar B$ pairs~\cite{a1pi_babar}.
It is illustrated in Figure~\ref{time_a1pi} by the projection plots onto $\Dt$
for $B^0$ and $\bzb$ tags, and the asymmetry between $B^0$ and $\bzb$ tags.
From $608 \pm 52$ signal events, it allows to measure
$\aeff^{\ap} = (79 \pm 7){\ensuremath{^\circ}}$.

To constrain $\alpha - \aeff$ and extract $\alpha$, flavor SU(3) symmetry
can be used~\cite{zupan}.
In addition to $\bap$, the related modes $B \to K_1 \pi$ and $B \to a_1 K$
have to be measured.
Using $383 \times 10^6~B\bar B$ pairs, \babar\ has studied 
the $B \to a_1 K$ modes~\cite{a1k_babar} 
and has measured the branching fractions 
${\cal B}(B^0 \to a_1^- K^+) = (16.3 \pm 2.9 \pm 2.3) 10^{-6}$ and
${\cal B}(B^+ \to a_1^+ K^0) = (34.9 \pm 5.0 \pm 4.4) 10^{-6}$.
The $B \to K_1 \pi$ modes, 
where $K_1$ is a mixture of $K_1(1270)$ and $K_1(1400)$, 
are being searched for.
With this last piece of information, a new constraint on $\alpha$ 
from the $\bap$ modes will be set.

\section{Summary}

\begin{figure}[htb]
\centering
\includegraphics[width=80mm]{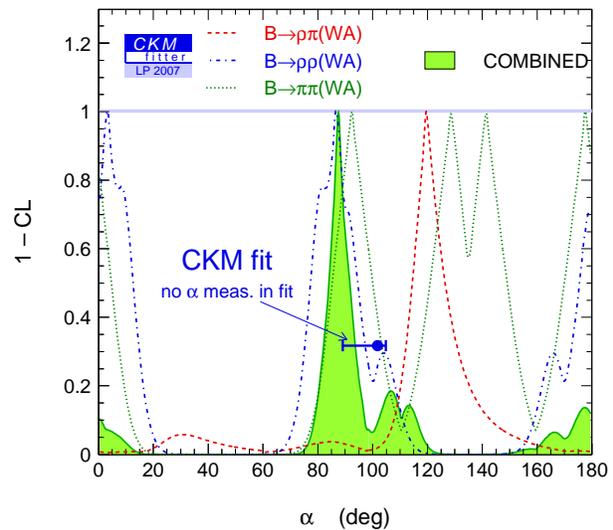}
\caption{Combined constraint on $\alpha$.
The individual constraints from the $\bpp$, $\brr$, and $\brp$ modes are shown
by the dotted, dot-dashed, and dashed curves respectively,
while the hatched region gives the combined constraint.}
\label{alpha_comb}
\end{figure}

Figure~\ref{alpha_comb} summarizes the constraints on $\alpha$ obtained 
by \babar\ and Belle, from the isospin analysis in the $\bpp$ and $\brr$ modes
and the Dalitz analysis in the $\brp$ modes.
The contribution from the $\bpp$ modes is limited
by the large penguin pollution.
The $\brr$ mode, slightly improved by the time-dependent $CP$ analysis of the 
$\brzrz$ channel, gives the best single measurement, 
but has mirror solutions that are disfavored thanks to the Dalitz $\brp$ 
analysis results.
The expected measurement of $\alpha$ in the $\bap$ modes will bring additional
information.

All the current results yield a combined value
of $\alpha = (87.5 ^{+6.2} _{-5.3})^\circ$.
This direct measurement of $\alpha$ is in good agreement with 
the indirect measurement $(102 ^{+3} _{-13})^\circ$ 
from the global CKM fit.

\bigskip 

\end{document}